\documentclass[graybox]{svmult}

\usepackage{mathptmx}       
\usepackage{helvet}         
\usepackage{courier}        
\usepackage{type1cm}        
\usepackage{makeidx}         
\usepackage{graphicx}        
\usepackage{multicol}        
\usepackage[bottom]{footmisc}
\usepackage{color}
\usepackage[justification=centering]{caption}
\usepackage[boxed,ruled,lined,linesnumbered]{algorithm2e}
\usepackage{ifpdf}
\usepackage{cite}

\usepackage{listings}
 \definecolor{dkgreen}{rgb}{0,0.6,0}
 \definecolor{mauve}{rgb}{0.58,0,0.82}
 \definecolor{lightyellow}{rgb}{1,1,.88}
 \definecolor{bronze}{rgb}{0.8, 0.5, 0.2}
 \definecolor{brown}{rgb}{0.59, 0.29, 0.0}
\lstdefinelanguage{MyR}{
  breaklines = true, 
  basicstyle=\color{brown}\footnotesize\ttfamily, 
  keepspaces=true,
  breakatwhitespace=false, 
  backgroundcolor=\color{lightyellow},
  commentstyle=\color{dkgreen},
  stringstyle=\color{mauve},
  }

\usepackage[cmex10]{amsmath}
\usepackage{subfigure}
\usepackage{algorithmic}
\usepackage{array}
\usepackage{float}
\usepackage{color}
\usepackage{multirow}
\usepackage{amsmath,amsfonts}
\usepackage[colorlinks]{hyperref}
\usepackage[boxed,ruled,lined,linesnumbered]{algorithm2e}
\usepackage{amssymb,amstext}
\usepackage{mdwmath}
\usepackage{mdwtab}
\usepackage{amsmath,amssymb}
\usepackage{eqparbox}
\usepackage{stfloats}
\usepackage{url}
\usepackage{setspace}
\usepackage{animate}

\def\t0{{t_0}}

\def\E{{\mathbb E}}

\def\N{{\mathbb N}}
\def\R{{\mathbb R}}        

\def\t0{{t_0}}

\def\NN{{\mathcal N}}

\def\prox{{\rm prox}}

\newcommand{\Rcommand}[1]{{\tt{\color{brown}{#1}}}}

\makeindex             

\begin{document}

\title{A Tutorial on $\Rcommand{Libra}$: R package for the Linearized Bregman Algorithm in High Dimensional Statistics}
\titlerunning{A Tutorial on ${\color{brown}{\tt Libra}}$}
\author{Jiechao Xiong, Feng Ruan, and Yuan Yao}
\institute{Jiechao Xiong \at Peking University, School of Mathematical Sciences, Beijing, China 100871, \email{xiongjiechao@pku.edu.cn}
\and Feng Ruan \at Stanford University, Department of Statistics, Sequoia Hall, Stanford, CA 94305, \email{fengruan@stanford.edu}
\and Yuan Yao \at Peking University, School of Mathematical Sciences, Beijing, China 100871, \email{yuany@math.pku.edu.cn}}
%
%
\maketitle


\abstract{The R package, $\Rcommand{Libra}$, stands for the LInearized BRegman Algorithm in high dimensional statistics. The Linearized Bregman Algorithm is a simple iterative procedure to generate sparse regularization paths of model estimation, which are firstly discovered in applied mathematics for image restoration and particularly suitable for parallel implementation in large scale problems. The limit of such an algorithm is a sparsity-restricted gradient descent flow, called the Inverse Scale Space, evolving along a parsimonious path of sparse models from the null model to overfitting ones. In sparse linear regression, the dynamics with early stopping regularization can provably meet the unbiased Oracle estimator under nearly the same condition as LASSO, while the latter is biased. Despite their successful applications, statistical consistency theory of such dynamical algorithms remains largely open except for some recent progress on linear regression. In this tutorial, algorithmic implementations in the package are discussed for several widely used sparse models in statistics, including linear regression, logistic regression, and several graphical models (Gaussian, Ising, and Potts). Besides the simulation examples, various application cases are demonstrated, with real world datasets from diabetes, publications of COPSS award winners, as well as social networks of two Chinese classic novels, Journey to the West and Dream of the Red Chamber.}

\section{Introduction to ${\color{brown}{\tt{Libra}}}$}
The free R package, ${\color{brown}{\tt Libra}}$, has its name as the acronym for the LInearized BRegman Algorithm (also known as Linearized Bregman Iteration in literature) in high dimensional statistics. It can be downloaded at 
\begin{center}
\url{https://cran.r-project.org/web/packages/Libra/index.html}
\end{center}

A parsimonious model selection with sparse parameter estimation has been a central topic in high dimensional statistics in the past two decades. For example, the following models are included in the package: 
\begin{itemize}
\item sparse linear regression,
\item sparse logistic regression (binomial, multinomial),
\item sparse graphical models (Gaussian, Ising, Potts). 
\end{itemize}
A wide spreading traditional approach is based on optimization to look for penalized M-estimators, i.e.
\begin{equation} 
\min_{\theta} L(\theta) + \lambda P(\theta), \ \ \ L(\theta):=\frac{1}{n}\sum_{i=1}^n l((x_i,y_i),\theta), 
\end{equation}
where $l((x_i,y_i),\theta)$ measures the loss of $\theta$ at sample $(x_i,y_i)$ and $P(\theta)$ is a sparsity-enforced penalty function on $\theta$ such as the $l_1$-penalty in LASSO \cite{lasso} and the nonconvex SCAD \cite{FanLi01}, etc. However, there are several shortcomings known in this approach: a convex penalty function will introduce bias to the estimators, while a nonconvex penalty, which may reduce the bias, yet suffers the computational hurdle to locate the global optimizer. Moreover, in practice a regularization path is desired which needs to search many optimizers $\theta_\lambda$ over a grid of regularization parameters $\{\lambda_j\geq 0: j\in \N\}$. 

In contrast, the Linearized Bregman (Iteration) Algorithm implemented in {\color{brown}{\tt Libra}} is based on the following iterative dynamics:
\begin{subequations}\label{eq:lbi0}
\begin{align}
\rho^{k+1} + \frac{1}{\kappa} \theta^{k+1} - \rho^k - \frac{1}{\kappa} \theta^k &=- \alpha_k \nabla_\theta L(\theta^k), \label{eq:lbi0-a}\\
 \rho^k &\in \partial P(\theta^k), \label{eq:lbi0-b}
\end{align}
\end{subequations}
with parameters $\alpha_k, \kappa>0$, and initial choice $\theta^0=\rho^0=0$. The second constraint requires that $\rho^k$ must be a subgradient of the penalty function $P$ at $\theta^k$. The iteration above can be restated in the following equivalent format with the aid of proximal map,  
\begin{subequations}\label{eq:lbi1}
\begin{align}
z^{k+1} & = z^k - \alpha_t \nabla_\theta L(\theta^k), \label{eq:lbi0-a}\\
 \theta^{k+1} &=\kappa \cdot {\prox}_{P}(z^{k+1}), \label{eq:lbi1-b}
\end{align}
\end{subequations}
where the proximal map associated with the penalty function $P$ is given by 
\[ \prox_P(z) = \arg\min_u \left( \frac{1}{2}\| u - z\|^2 + P(z) \right ). \]

The Linearized Bregman Iteration \eqref{eq:lbi0} generates a parsimonious path of sparse estimators, $\theta^t$, starting from a null model and evolving into dense models with different levels of sparsity until reaching overfitting ones. Therefore the dynamics itself can be viewed as regularization paths. Such an iterative algorithm was firstly introduced in \cite{YODG08} (Section 5.3, Equations (5.19) and (5.20)) as a scalable algorithm for large scale problems of image restoration with TV-regularization and compressed sensing, etc. As $\kappa\to \infty$ and $\alpha_t\to 0$, the iteration has a limit dynamics, known as Inverse Scale Space (ISS) \cite{iss} describing its evolution direction from the null model to full ones, 
\begin{subequations}\label{eq:iss}
\begin{align}
\frac{d \rho(t)}{d t} &=- \nabla_\theta L(\theta(t)), \label{eq:iss-a}\\
 \rho(t) &\in \partial P(\theta(t)). \label{eq:iss-b}
\end{align}
\end{subequations}
The computation of such ISS dynamics is discussed in \cite{Burger13aiss}. With the aid of ISS dynamics, recently \cite{osher2014} establishes the model selection consistency for early stopping regularization in both ISS and Linearized Bregman Iterations for the basic linear regression models. In particular, under nearly the same conditions as LASSO, ISS finds the oracle estimator which is bias-free while the LASSO is biased. However, it remains largely open to explore the statistical consistency for general loss and penalty functions, despite successful applications of \eqref{eq:lbi0} in a variety of fields such as image processing and statistical modeling that will be illustrated below. As one purpose of this tutorial, we hope more statisticians will benefit from the usage of this simple algorithm with the aid of this R package, {\color{brown}{\tt Libra}}, and eventually reach a deep understanding of its statistical nature.

In the sequel we shall consider two types of parameters, $(\theta_0,\theta)$, where $\theta_0$ denotes  
the unpenalized parameters (usually intercept in the model) and $\theta$ represents all the penalized sparse parameters. 
Correspondingly, $L(\theta_0,\theta)$ denotes the Loss function. In most cases, $L(\theta_0,\theta)$ is the same as the negative log-likelihood function of the model. 

Two types of sparsity-enforcement penalty functions will be studied here:  
\begin{itemize}
\item LASSO ($l_1$) penalty for entry-wise sparsity: 
$$ P(\theta)=\|\theta\|_1:=\sum_j |\theta_j|;$$ 
\item Group LASSO ($l_1$-$l_2$) penalty for group-wise sparsity: 
$$P(\theta)=\|\theta\|_{1,2}=\sum_g \|\theta_g\|_2:=\sum_{g} \sqrt{\sum_{j:g_j=g}\theta_j^2},$$
\end{itemize}
where we use $\mathcal{G} = \{g_j:g_j~\mbox{is~the~group~of}~\theta_j, j=1,2,\dots,p\}$ to denote a disjoint 
partition of the index set $\{1, 2, \ldots, p\}$--that is, each group $g_j$ is a subset of the index set. When $\mathcal{G}$ is degenerated, i.e, $g_j=j, j=1,2,\dots,p$, the Group Lasso penalty is the same as the LASSO penalty. The proximal map for Group LASSO penalty is given by 
\begin{equation}
\prox_{\|\theta\|_{1,2}} (z)_j := 
\left\{
\begin{array}{ll}
\left(1-\frac{1}{\sqrt{\sum_{i:g_i=g_j}z_i^2}}\right)z_j, & \|z_{g_j}\|_2 \geq 1, \\
0, & \mbox{otherwise},
\end{array}
\right.
\end{equation}
which is also called the ${\mathbf{Shrinkage}}$ operator in literature. 

When the entry-wise sparsity is enforced, the parameters to be estimated in the models are 
encouraged to be `sparse' and treated independently. On the other hand, when the group-wise sparsity is enforced, it not only encourages 
the estimated parameters to be sparse, but also expects variables within the same group 
to be either selected or not selected at the same time. Hence, the group-wise sparsity requires 
prior knowledge of the group information of the correlated variables.

Once the parameters $(\theta_0,\theta)$, the loss function and group vectors are specified, the Linearized Bregman Iteration algorithm in \eqref{eq:lbi0} or \eqref{eq:lbi1} can be 
adapted to the new setting with partial sparsity-enforcement on $\theta$, as shown in Algorithm \ref{alg-LBI}. 
The iterative dynamics computes a regularization path for the parameters at different levels of sparsity -- starting from the null model with $(\theta_0,0)$, it evolves along a path of sparse models into the dense ones minimizing the loss. 

\begin{algorithm}[!h]
{
\caption{Linearized Bregman Algorithm.}\label{alg-LBI}
\textbf{Input:} Loss function $L(\theta_0,\theta)$, group vector $\mathcal{G}$, damping factor $\kappa$, step size $\alpha$.\\
\textbf{Initialize:} $k=0, t^k=0,\theta^k=0, z^k = 0, \theta_0^k=\arg \min_{\theta_0} L(\theta_0,0)$.\\
{\textbf{for $k=1,\dots,K$ do}
\begin{itemize}
  \item $z^{k+1} = z^{k} - \alpha\nabla_{\theta}L(\theta_0^k,\theta^k)$.
  \item $\theta^{k+1} = \kappa \cdot \mathbf{Shrinkage}(z^{k+1},\mathcal{G})$.
  \item $\theta_0^{k+1} = \theta_0^{k} - \kappa\alpha\nabla_{\theta_0}L(\theta_0^k,\theta^k)$.
  \item $t^{k+1} = (k+1)\alpha$.
\end{itemize}
\textbf{end for}}\\
\textbf{Output:} Solution path $\{t^k, \theta_0^k,\theta^k\}_{k= 0,1,\dots,K}$.\newline
where $\theta = \mathbf{Shrinkage}(z,\mathcal{G})$ is defined as:~$\theta_j = \mathbf{max}\left(0,1-\frac{1}{\sqrt{\sum_{i:g_i=g_j}z_i^2}}\right)z_j$.
}
\end{algorithm}

In the following Section \ref{sec: linear model}, \ref{sec: logistic model}, and \ref{sec: graphical model}, we shall specialize such a general algorithm in linear regression, logistic regression, and graphical models, respectively. Section \ref{sec-detail} includes a discussion on some universal parameter choices. Application examples will be demonstrated along with source codes.

\section{Linear Model}
\label{sec: linear model}
In this section, we are going to show how the Linearized Bregman (LB) algorithm and the Inverse Scale Space (ISS) fit sparse 
linear regression model.  Suppose we have some covariates $x_i \in \R^p$ 
for $i=1,2, \ldots, n$. The responses $y_i$ with respect to $x_i$, where $i=1,2,\ldots, n$, 
are assumed to follow the linear model below: 
\begin{equation*}
y_i = \theta_0 + x_i^T \theta + \epsilon, \epsilon \sim \NN(0, \sigma^2).
\end{equation*}
Here, we allow the dimensionality of covariates $p$ to be either smaller or greater than the sample size $n$. 
Note that, in latter case, we need to make additional sparsity 
assumptions on $\theta$ in order to make the model identifiable (and also, make recovery of $\theta$ possible). 
Both the Linearized Bregman Algorithm and ISS compute their own `regularization paths' for the (sparse) linear model. 
The statistical properties for the two regularization paths for linear models 
are established in~\cite{osher2014} where the authors show that under some natural conditions for both regularization paths, 
some points on the paths determined by a data-dependent
early-stopping rule can be nearly unbiased and exactly recover the support of signal $\theta$. Note that
the latter exact recovery of signal support can have a significant meaning in the regime where $p \gg n$, in 
which case, an exact variable selection work is done simultaneously with the model fitting process.  
In addition, the computational cost for regularization path generated by LB algorithm is relatively cheap
in linear regression model case, compared to many other existing methods. We refer the readers to~\cite{osher2014} for more details. 
Owning both statistical and computational advantages over other methods, 
the Linearized Bregman Algorithm is strongly recommended for practitioners, 
especially for those who are dealing with computationally heavy tasks. 

Here, we give a more detailed illustration on how the Linearized Bregman Algorithm computes the solution 
path for the linear model. We use negative log-likelihood as our loss function, 
\begin{equation*}
L(\theta_0,\theta) = \frac{1}{2n}\sum_{i=1}^n (y_i-\theta_0-x_i^T\theta)^2.
\end{equation*}
To compute the regularization path, we need to compute the gradient of loss with
respect to its parameters $\theta_0$ and $\theta$, as is shown in Algorithm~\ref{alg-LBI},
\begin{eqnarray}
\nabla_{\theta_0}L(\theta_0,\theta) &=\frac{1}{n}\sum_{i=1}^n -(y_i-\theta_0-x_i^T\theta),\nonumber\\
\nabla_{\theta}L(\theta_0,\theta) &=\frac{1}{n}\sum_{i=1}^n -x_i(y_i-\theta_0-x_i^T\theta).\nonumber
\end{eqnarray}
In linear model, each iteration of the Linearized Bregman Algorithm requires $O(np)$ FLOPs 
in general (and the cost can be cheaper if additional sparsity structure on parameters are known), and 
the overall time complexity for the entire regularization path is $O(npk)$, where $k$ is the number of 
iterations. The number of iterations in the Linearized Bregman Algorithm is dependent on the 
underlying step-size $\alpha$, which can be understood as the counterpart of learning rate 
that appear in the standard gradient descent algorithms. For practitioners, choosing parameters
$\alpha$ needs a deeper understanding of the standard tradeoffs between statistical and computational issues here. 
With a high learning rate $\alpha$, the Linearized Bregman Algorithm can generate a `coarse' 
regularization path in only a few iterations. Yet such `solution' path might not be statistically informative; 
with only a few points on the path, practitioners may not be able to determine which of these points
actually recover the true support of the unknown signal $\theta$. On the other hand, a `denser' 
solution path generated by low learning rate $\alpha$ provide more information about the
true signal $\theta$, yet it might lose some computational efficiency of the algorithm itself. 

In addition to the parameter $\alpha$, another parameter $\kappa$ is needed in the algorithm. As $\kappa\to \infty$ and $\alpha\to 0$, the Linearized Bregman Algorithm \eqref{eq:lbi0}
will converge to its limit ISS \eqref{eq:iss}. Therefore, with a higher value of $\kappa$, the Linearized Bregman Algorithm will have a stronger effect on 
`debiasing' the path, and hence give a better estimate of the underlying signal at a cost of possible high variance. Moreover, the parameters $\alpha$ and $\kappa$ need to satisfy 
\begin{equation} \label{eq:alphakappa}
\alpha \kappa \| S_n \| \le 2, \ \ \ S_n=\frac{1}{n} \sum_{i=1}^n x_i x_i^T, 
\end{equation}
otherwise the Linearized Bregman iterations might oscillate and suffer numerical convergence issues~\cite{osher2014}. Therefore in practice, one typically first chooses $\kappa$ which might be large enough, then follows a large enough $\alpha$ according to \eqref{eq:alphakappa}. In this sense, $\kappa$ is the essential free parameter.  

Having known how the Linearized Bregman Algorithm work in linear model, we are ready to introduce the 
command in $\Rcommand{Libra}$ that can be used to generate the path, 
\begin{equation*}
\small{\Rcommand{lb(X,y,kappa,alpha,tlist,family=``gaussian",group=FALSE,index=NA)}}
\end{equation*}
In using the command above, the user \emph{must} give inputs for the design matrix $\Rcommand{X}\in \R^{n\times p}$, 
the response vector $\Rcommand{y}\in \R^n$ and the parameter $\Rcommand{kappa}$. Notably, the parameter $\Rcommand{alpha}$ 
is not required to be given in the use of such command, and in the case when it's missing, an 
internal value for $\Rcommand{alpha}$ satisfying \eqref{eq:alphakappa} would be used and this internally-generated $\Rcommand{alpha}$ would guarantee the 
convergence of the algorithm. The $\Rcommand{tlist}$ is a group of parameters $t$ that determine the output of the above command. 
When the ${\tt{\color{brown}{tlist}}}$ is given, only points at the pre-decided set of ${\tt{\color{brown}{tlist}}}$ on the regularization path will be returned. 
When it is missing, then a data dependent ${\tt{\color{brown}{tlist}}}$ will be calculated. See Section~\ref{sec-detail} for more details 
on the ${\tt{\color{brown}{tlist}}}$. Finally, when group sparsity is considered, the user needs to input an additional argument ${\tt{\color{brown}{index}}}$
to the algorithm so that it can know the group information on the covariates. 

As the limit of Linearized Bregman iterations when $\kappa\rightarrow\infty,\alpha\rightarrow 0$, 
the Inverse Scale Space for linear model with $l_1$-penalty is also available in our $\Rcommand{Libra}$ package:
\begin{equation*}
\Rcommand{iss(X, y, intercept = TRUE, normalize = TRUE)}.
\end{equation*}
As is suggested by the previous discussion on the effect of $\kappa$ on the regularization path, 
the ISS has the strongest power of `debiasing' the path; once the model selection consistency is reached, it can return the `oracle' unbiased estimator! 
Yet one disadvantage of ISS solution path is its relative computational inefficiency compared to the Linearized 
Bregman Algorithm. 

\subsection{Example: Simulation Data}
\begin{figure}[!h]
\centering
\includegraphics[width=0.7\textwidth]{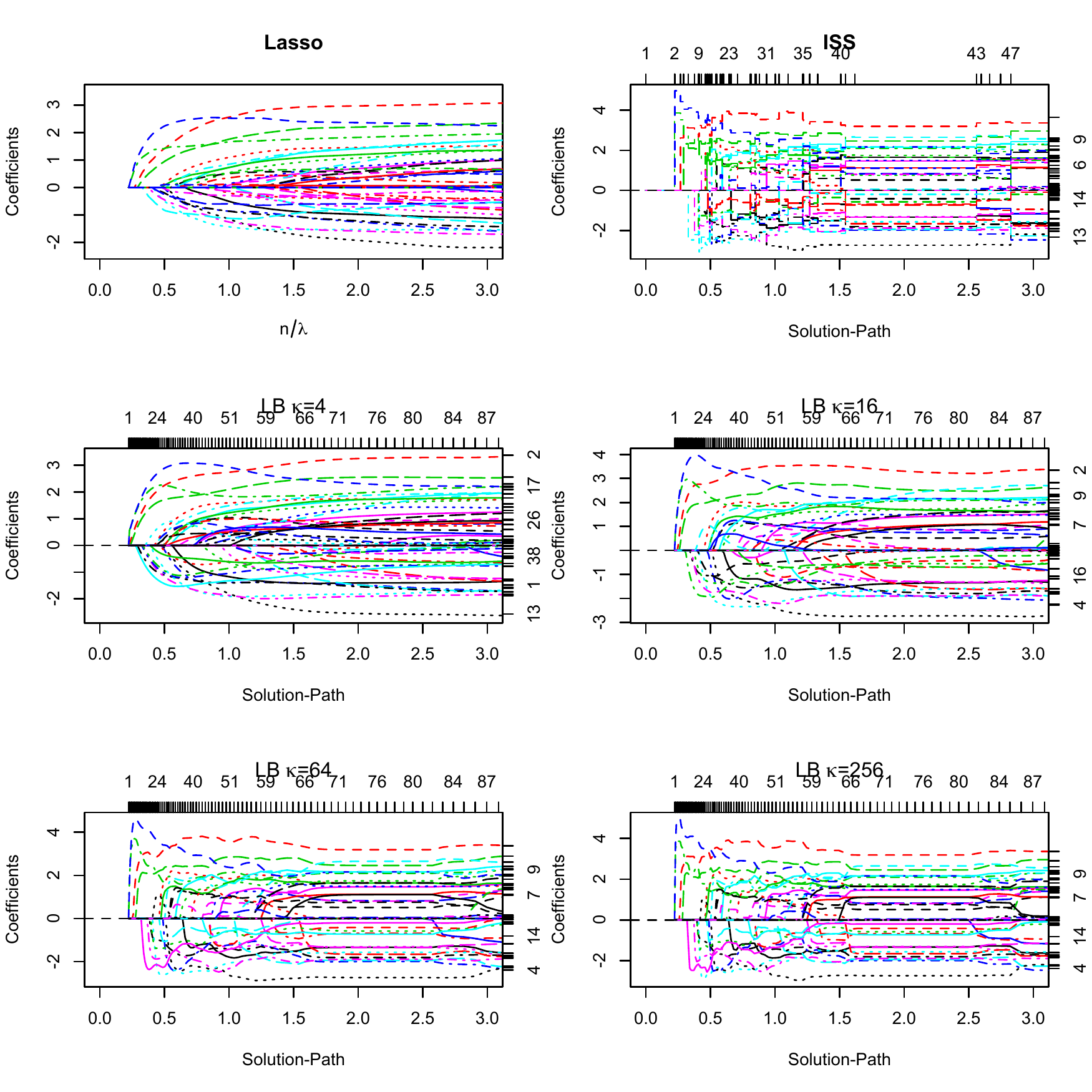}
\caption{Regularization paths of LASSO, ISS, and LB with different choices of $\kappa$ ($\kappa=2^2, 2^4, 2^6, 2^8$, and $\alpha\kappa$ = 1/10). As $\kappa$ grows, the paths of Linearized Bregman iterations approach that of ISS. The x-axis is $t$.} \label{fig:osher}
\end{figure}

Here is the example in \cite{osher2014}. A comparison of regularization paths generated by LASSO, ISS and the Linearized Bregman iterations is shown in Figure \ref{fig:osher}.

\begin{lstlisting}[language=MyR,caption={}]
library(MASS)
library(lars)
library(Libra)
n = 80;p = 100;k = 30;sigma = 1
Sigma = 1/(3*p)*matrix(rep(1,p^2),p,p)
diag(Sigma) = 1
A = mvrnorm(n, rep(0, p), Sigma)
u_ref = rep(0,p)
supp_ref = 1:k
u_ref[supp_ref] = rnorm(k)
u_ref[supp_ref] = u_ref[supp_ref]+sign(u_ref[supp_ref])
b = as.vector(A%*%u_ref + sigma*rnorm(n))
lasso = lars(A,b,normalize=FALSE,intercept=FALSE,max.steps=100)
par(mfrow=c(3,2))
matplot(n/lasso$lambda, lasso$beta[1:100,], xlab = bquote(n/lambda), 
  ylab = "Coefficients", xlim=c(0,3),ylim=c(range(lasso$beta)),type='l', main="Lasso")
object = iss(A,b,intercept=FALSE,normalize=FALSE)
plot(object,xlim=c(0,3),main=bquote("ISS"))
kappa_list = c(4,16,64,256)
alpha_list = 1/10/kappa_list
for (i in 1:4){
  object <- lb(A,b,kappa_list[i],alpha_list[i],family="gaussian",group=FALSE,
               trate=20,intercept=FALSE,normalize=FALSE)
  plot(object,xlim=c(0,3),main=bquote(paste("LB ",kappa,"=",.(kappa_list[i]))))
}
\end{lstlisting}

\subsection{Example: Diabetes Data}
\begin{figure}[!h]
\centering
\includegraphics[width=0.7\textwidth]{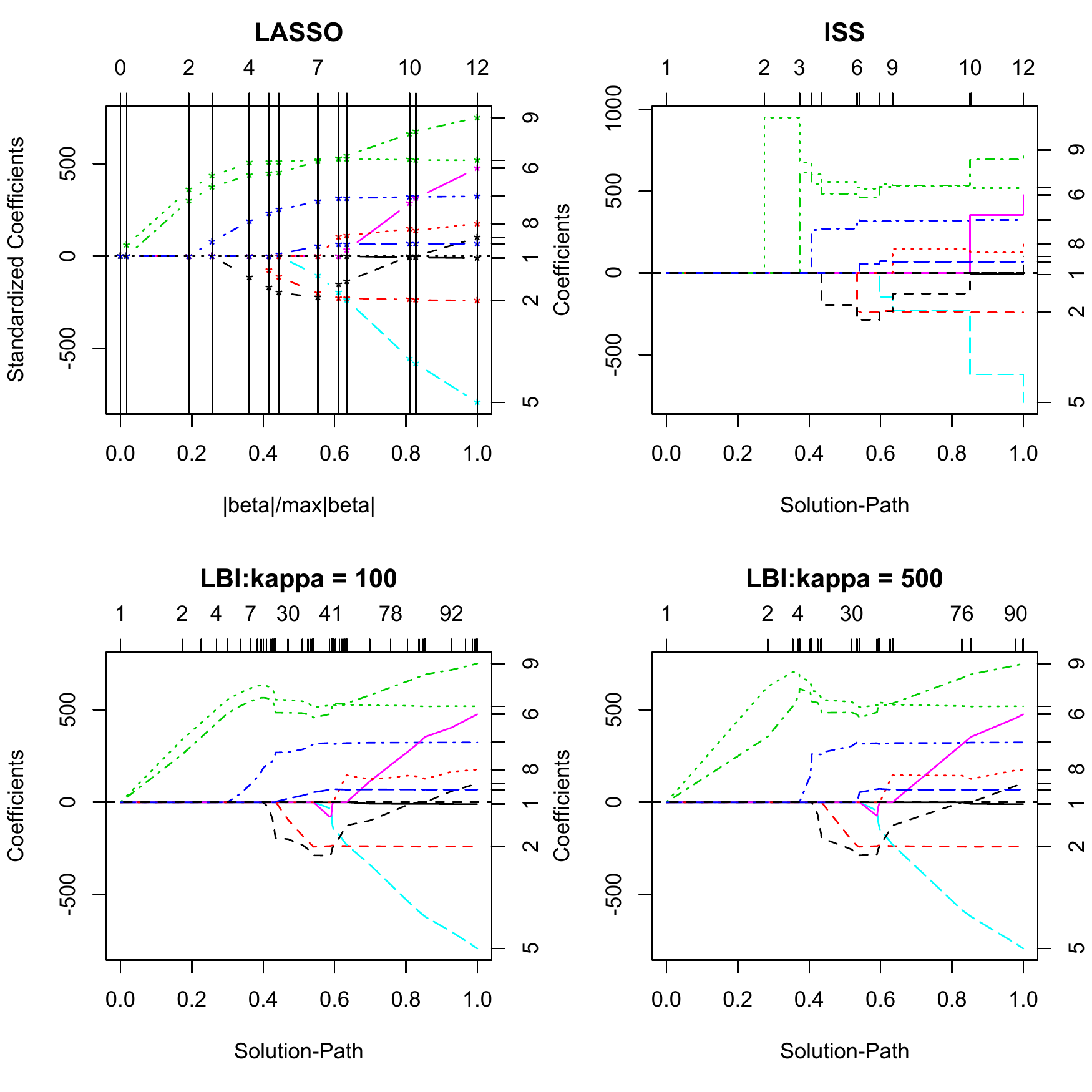}
\caption{Regularization paths of LASSO, ISS, and Linearized Bregman Iterations on diabetes data. The piecewise constant regularization path of ISS exhibits strong debiasing effect. The orders of variables entering into nonzero regimes are similar in different paths. The x-axis is $\|\theta\|_1$.} \label{fig:diabetes}
\end{figure}

A diabetes dataset is used as an example in~\cite{lars} to illustrate the $\Rcommand{lars}$ algorithm. The dataset contains 442 samples (diabetes patients) with 10 baseline variables. Here, we show the solution paths of both the Linearized Bregman Algorithm and ISS on the data, assuming a sparse linear regression model 
between the baseline variables and the response. The LASSO regularization path is computed by R-package $\Rcommand{lars}$. Figure \ref{fig:diabetes} shows the comparison of different paths. It can be seen that the LASSO path is continuous, while the ISS path is piece-wise constant exhibiting the strong `debiasing' effect. The paths generated by discrete Linearized Bregman iterations somehow lie between them. It is easy to see the sudden `shocks' in the figure when the variables are picked up in the regularization path of the ISS or in the paths of Linearized Bregman iterations with large $\kappa$. These `shocks' correspond to the stronger debiasing effect of the Linearized Bregman Algorithm and ISS than LASSO. Hence our signals can be fitted in a more `aggressive' way than ${\color{brown}{\tt lars}}$ when we use a strong regularization. Although the curve shapes of these paths are different, it is noticeable that the order of those paths entering into nonzero regimes bears a great similarity, which implies that the model selection effects of these algorithms are similar in this dataset. 

\begin{lstlisting}[language=MyR,caption={}]
library(lars)
library(Libra)
data(diabetes)
attach(diabetes)

lasso <- lars(x,y)
par(mfrow=c(2,2))
plot(lasso)

issobject <- iss(x,y)
plot(issobject,xtype="norm")  #plot.lb
title("ISS",line = 2.5)

kappa <- c(100,500)
for (i in 1:2){
    object <- lb(x,y,kappa[i],family="gaussian",trate=1000)
    plot(object,xtype="norm")
    title(paste("LBI:kappa =",kappa[i]),line = 2.5)
}
detach(lasso)
\end{lstlisting}

\section{Logistic Model}
\label{sec: logistic model}
\subsection{Binomial Logistic Model}
In binary classification task, one of the mostly widely used model is the binomial logistic model, see \cite{esl}.
Given the i.i.d data $(x_i, y_i) \in \R^p \times  \{\pm 1\}$, the standard binomial logistic model assumes the following 
predictive relationship between the covariates $x_i \in \R^p$ and their response $y_i \in \{\pm 1\}$ for $i=1, 2, \ldots, n$: 
\begin{equation*}
\frac{P(y_i=1|x_i)}{P(y_i=-1|x_i)} = \exp(\theta_0 + x_i^T\theta),
\end{equation*}
where, in the above equation, 
$\theta\in \R^p$ represents the regression coefficients before the covariates and $\theta_0 \in \R$ represents the offset 
in the regression model. Here, we allow the dimensionality $p$ to be greater than or equal to the sample size $n$. As is discussed 
in the linear regression case, when $p > n$, additional sparsity assumptions on the regression coefficient $\theta$ should be enforced 
to make the logistic model identifiable from the data (and also, recovery of the parameters $\theta$ possible). 
The goal of this section is to show how the Linearized Bregman Algorithm fits the sparse binomial logistic regression model 
in high dimension. An early version of the Linearized Bregman iterations was implemented in \cite{ShiYinOsh13}, which differs to Algorithm \ref{alg-LBI} mainly in their zero initialization where we exploit an optimal choice of $\theta_0$ as a maximum likelihood estimate restricted to the null sparse model $\theta=0$. See more discussions on initializations in Section \ref{sec-detail}.

As is discussed similarly in the linear regression case, a regularization path is returned via the Linearized Bregman Algorithm, 
where practitioners can find different estimates of the same parameters under different level of sparsity assumptions 
on the true parameter $\theta$. To give a more detailed illustration on how the Linearized Bregman Algorithm 
computes the regularization path, we first introduce the loss function in the algorithm, which is given by the negative log-likelihood 
of the binomial model: 
\begin{equation*}
L(\theta_0,\theta) = \frac{1}{n}\sum_{i=1}^n \log(1+\exp(-y_i(\theta_0 + x_i^T\theta))).
\end{equation*}
To compute the regularization path, the Linearized Bregman Algorithm~\ref{alg-LBI} needs to evaluate the derivatives of the loss function with 
respect to $\theta$ and $\theta_0$ for each of the iteration point in the path, 
\begin{eqnarray}
\nabla_{\theta_0}L(\theta_0,\theta) &=&\frac{1}{n}\sum_{i=1}^n \frac{-y_i}{1+\exp(y_i(\theta_0 + x_i^T\theta))},\nonumber\\
\nabla_{\theta}L(\theta_0,\theta) &=&\frac{1}{n}\sum_{i=1}^n \frac{-y_ix_i}{1+\exp(y_i(\theta_0 + x_i^T\theta))}.\nonumber
\end{eqnarray}
In binomial logistic model, each iteration of the Linearized Bregman Algorithm requires $O(np)$ FLOPS in general, and the 
overall time complexity for the entire solution path is $O(npk)$, where $k$ is the number of iterations. 

Here, we give the command in $\Rcommand{Libra}$ that can be used to generate the path for the logistic model, 
\begin{equation*}
{\color{brown}{\tt lb(X,y,kappa,alpha,tlist,family=``binomial",group=FALSE,index=NA)}}.
\end{equation*}
As is shown in the above command, the user is required to provide data $\Rcommand{X}$, $\Rcommand{y}$, as well as the parameters $\Rcommand{alpha}$, $\Rcommand{kappa}$,
and $\Rcommand{tlist}$. The effects of these parameters on the resulting regularization paths for binomial logistic model 
parallel that for the linear model. Hence, we refer the reader to section~\ref{sec: 
linear model} to find a detailed explanation on how the parameters affect the regularization paths. Finally, similar to the case in linear regression, 
if one needs to enforce a particular group sparse structure on the output parameters $\theta$, he/she has to input the
$\Rcommand{index}$ argument so that the algorithm can know the group information assumption on the covariates.  

\subsubsection{Example: Publications of COPSS Award Winners} 
\begin{figure}[!h]
\begin{center}
\includegraphics[width=0.5\textwidth]{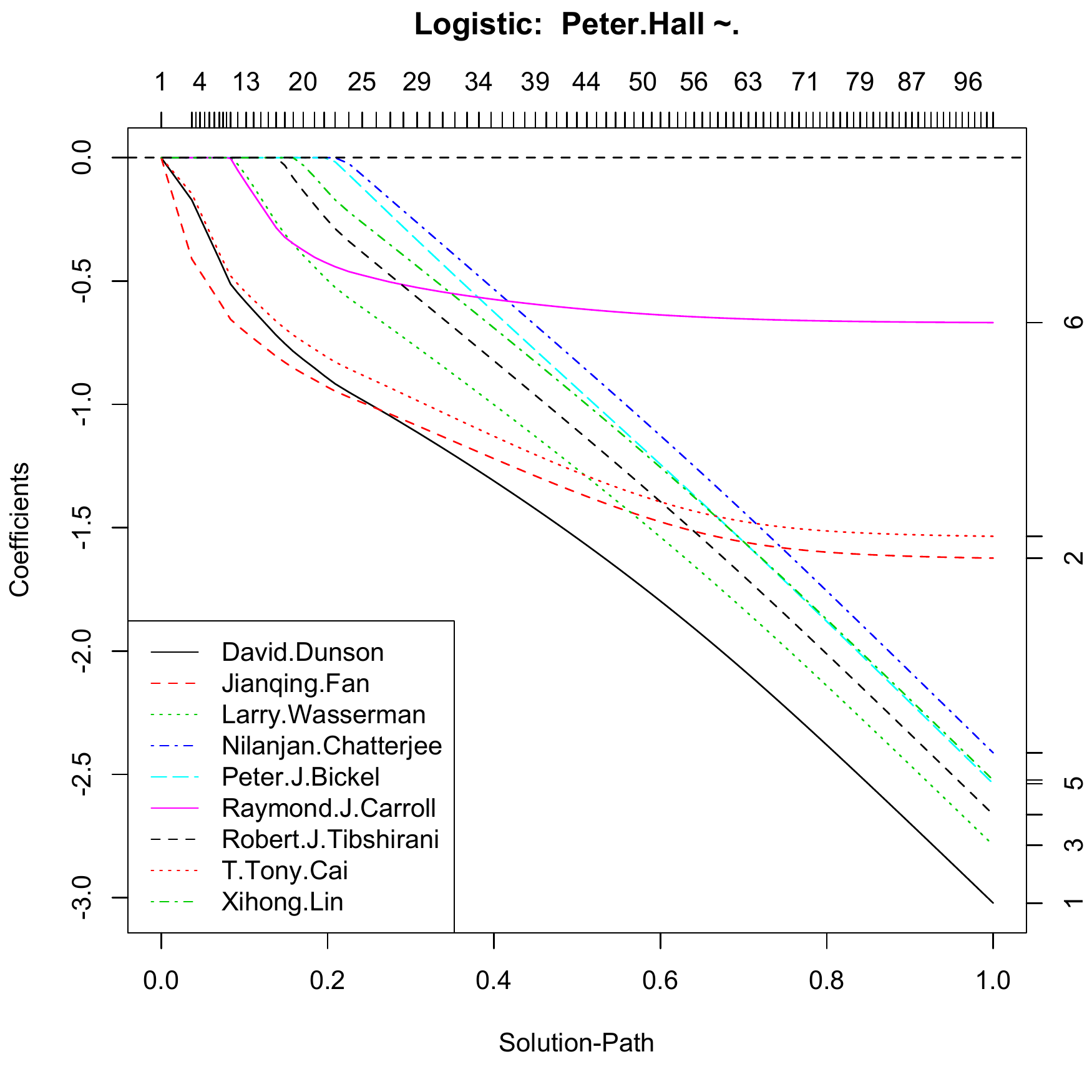} 
\end{center}
\caption{Regularization path of logistic regression by LB on COPSS data. The $x$-axis is normalized $\theta\|_1$. As all the coefficients on the paths appear to be negative, it suggests that the probability of these COPSS award winners collaborating with Peter Hall is below the average in a contrast to his fruitful publications. The three paths which level off as $\|\theta\|_1$ grows correspond to Jianqing Fan, Tony Cai, and Raymond J Carroll, who are the only collaborators of Peter Hall in this dataset.}
\label{fig:copss10}
\end{figure}

The following example explores a statistician publication dataset provided by Professor Jiashun Jin at Carnegie Mellon University \cite{copss}. The dataset consists of 3248 papers by 3607 authors between 2003 and the first quarter of 2012 from the following four journals: the Annals of Statistics, Journal of the American Statistical Association, Biometrika and Journal of the Royal Statistical Society Series B. Here we extract a subset of 382 papers co-authored by 35 COPSS award winners. Peter Gavin Hall (20 November 1951 – 9 January 2016) is known as one of the most productive statisticians in history and contributed 82 papers in this dataset. Can we predict the probability of his collaborations with other COPSS award winners? A logistic regression model will be used for this exploration. For a better visualization, we only choose 9 other COPSS winners who have no less than 10 papers in this dataset. The following codes compute regularization paths of the Linearized Bregman iterations for logistic regression model to predict the probability of Peter Hall's collaborations with them. From the regularization paths shown in Figure \ref{fig:copss10}, it can be seen that the probability of collaborations between Peter Hall and other COPSS winners are all reduced below the average indicated by the negative coefficients, which suggests that these COPSS winners usually work independently even occasionally coauthor some papers. The three paths which level off as iterations go correspond to Jianqing Fan, Tony Cai, and Raymond J Carroll, who are the only collaborators of Peter Hall in this dataset.
\begin{lstlisting}[language=MyR,deletekeywords={beta,kappa,list,path,names},caption={}]
library(Libra)
data<-read.table("copss.txt")
s0<-colSums(data)
data1<-data[,s0>=10]	# choose the authors whose publications are of no less than 10

y<-as.vector(2*as.matrix(data1[,5])-1); # Peter.Hall as response
X<-as.matrix(2*as.matrix(data1[,-5])-1); # Other COPSS winners as predictors
path <- lb(X,y,kappa = 1,family="binomial",trate=100,normalize = FALSE)

plot(path,xtype="norm",omit.zeros=FALSE)
title(main=paste("Logistic: ",attributes(data1)$names[5],"~."),line=3)
legend("bottomleft", legend=attributes(data1)$names[-5], col=c(1:6,1:3),lty=c(1:5,1:4))
\end{lstlisting}

\subsubsection{Example: Journey to the West}
\label{sec: journey to the west}

Journey to the West is one of the Four Great Classical Novels of Chinese Literature. The literature describes an adventure story 
about $\Rcommand{Tangseng}$ who travelled to the `West Regions' for Sacred Texts. The literature contains more than a hundred chapters and involves 
more than a thousand of characters. One interesting study on the literature would be to understand the social relationships between the main characters, 
i.e., to understand how those with different personalities and power can come along with each other. 

Here, we give a simple example showing how the Linearized Bregman Algorithm can be used to analyze the relationship between 
one main character, $\Rcommand{Monkey King~(Sunwukong)}$, to the other main characters. We collect some data that documents the appearance/disappearance of the top 
10 main characters under the pre-specified 408 different scenes in the novel. To analyze the relationship between $\Rcommand{Monkey King}$ to the 
other 9 main characters, we build up a logistic regression model, where the response $Y$ corresponds to the indicator of the appearance of the 
$\Rcommand{Monkey King}$ in these scenes and the other covariates $X$ correspond to the indicators of the appearance of the other 9 characters in the scenes. 
The data is collected via crowdsourcing at Peking University, and can be downloaded at the following course website 

\begin{center}
\url{http://math.stanford.edu/~yuany/course/2014.fall/}.
\end{center}

Below we analyze the result of the logistic regression model fitted by the Linearized Bregman Algorithm. 
Notice that, $\Rcommand{Tangseng}$, $\Rcommand{Pig~(Zhubajie)}$ and $\Rcommand{Friar Sand~(Shaseng)}$ are the first three main characters that are picked up in the regularization path.
In addition, the coefficients of their corresponding covariates are all positive, meaning that they probably show up the same time as the 
$\Rcommand{Monkey King}$ in the story. A combination of the above two phenomena is explained by the fact that in the novel they together with $\Rcommand{Monkey King~(Sunwukong)}$ form the fellowship of the journey to the west. On the other hand, $\Rcommand{Yuhuangdadi}$, $\Rcommand{Guanyinpusa}$ and $\Rcommand{Muzha}$ seem to have less involvements with the $\Rcommand{Monkey King}$, as 
they didn't show up in the paths until very late stages, with estimated coefficients being negative, indicating that they just appeared occasionally with the $\Rcommand{Monkey King}$ when he got troubles. 

\begin{figure}[!h]
\centering
\includegraphics[width=0.5\textwidth]{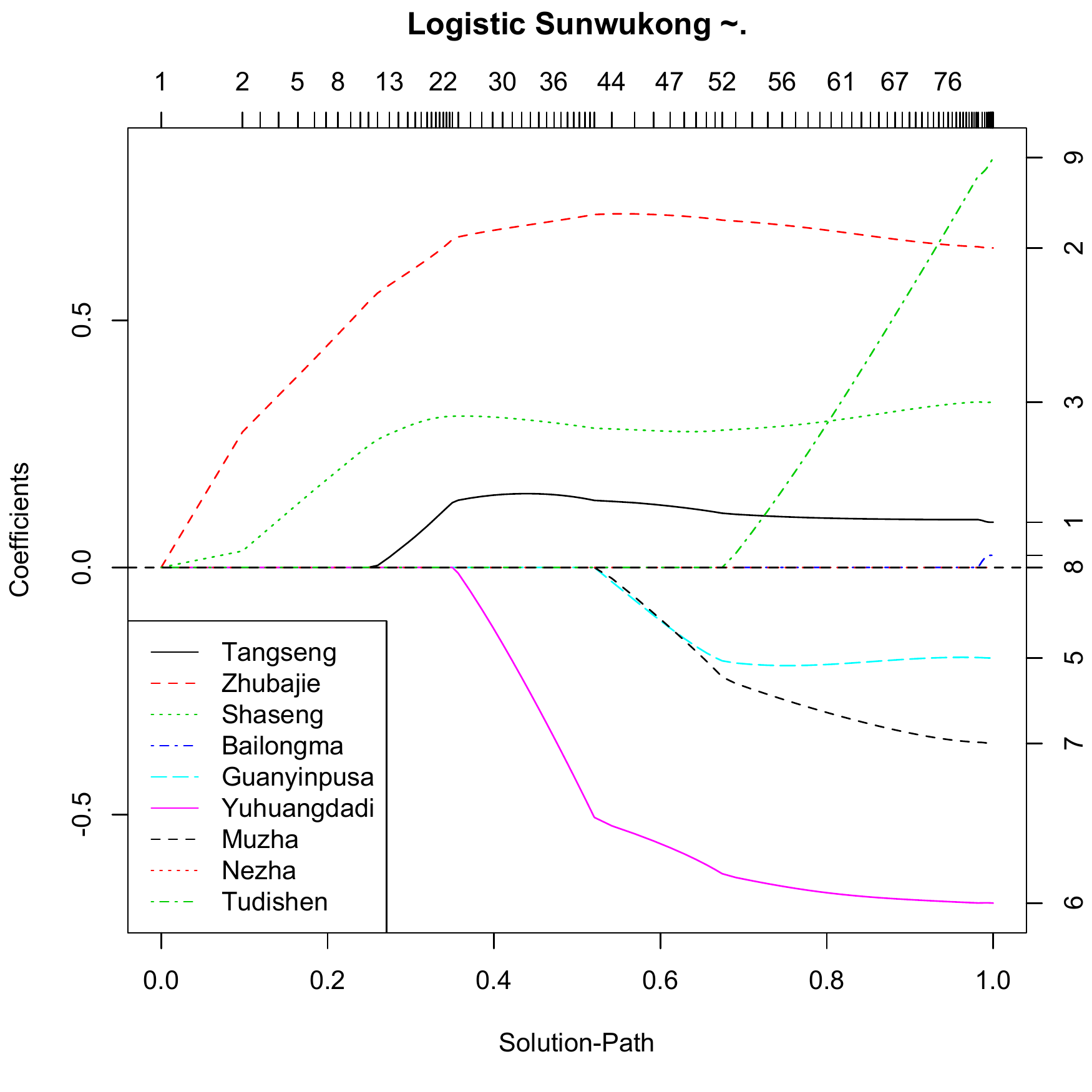}
\caption{Regularization path of $\Rcommand{lb}$ on $\Rcommand{west10}$ data using $\Rcommand{family="binomial"}$. The fellowship of the journey to the west is formed by $\Rcommand{Sunwukong}$ ($\Rcommand{MonkeyKing}$) and his three peers: $\Rcommand{Tangseng}$, $\Rcommand{Zhubajie}$, and $\Rcommand{Shaseng}$, corresponding to the first three paths.}
\end{figure}

\begin{lstlisting}[language=MyR,deletekeywords={beta,kappa,list,path,names},caption={}]
library(Libra)
data(west10)
y<-2*west10[,1]-1;
X<-as.matrix(2*west10[,2:10]-1);

path <- lb(X,y,kappa = 1,family="binomial",trate=100,normalize = FALSE)
plot(path,xtype="norm",omit.zeros=FALSE)
title(main=paste("Logistic",attributes(west10)$names[1],"~."),line=3)
legend("bottomleft", legend=attributes(west10)$names[-1], col=c(1:6,1:3),lty=c(1:5,1:4))
\end{lstlisting}

\newcommand{\defeq}{:=}
\subsection{Multinomial Logistic Model}

Multinomial logistic regression is a method that generalizes the binary logistic model to multi-class classification problems, 
where the response $y$ has $K(\geq 2)$ different outcomes \cite{esl}. The model assumes the 
following relationship between the response $y \in \{1, 2, \ldots, K\}$ and its covariate $x \in \R^p$: 
\begin{equation*}
P(y=k|x) = \frac{\exp(\theta_{k0} + x^T\theta_{k})}{\sum_{k=1}^{K}\exp(\theta_{k0} + x^T\theta_{k})}
\end{equation*}

As is discussed in the previous sections, often additional sparsity assumptions on the coefficients 
$\theta_k$ for $k=1, 2, \ldots, K$ are added by researchers to make the model more identifiable/more interpretable 
in high dimensions. Usually, researchers can have different prior beliefs on the underlying sparse structure of the model, 
and these different types of sparse structures correspond to different types of sparse multinomial logistic regression model. 
In our package, we consider three major variants of the original multinomial logistic model,
 i.e., the \emph{entry-wise} sparse, the \emph{column-wise} sparse and the \emph{block-wise} sparse multinomial logistic model. 
The entry-wise sparse model corresponds to adding an LASSO ($l_1$) penalty on all the parameters $\theta_k$ for $k=1,2,\ldots, K$. 
The column-wise sparsity corresponds to adding a more complicated group LASSO penalty on each column group of parameters $\theta_k$, $\sum_{j=1}^p\sqrt{\sum_{k=1}^K \theta_{kj}^2}$. Since each column of $\theta$ correspond to a feature $x_i$ for some $1\leq i\leq p$, getting column-wise sparse estimates
will select the same set of features for different response classes simultaneously. 
Finally, as a generalization of the previous group sparse model,
the block-wise sparse model assumes an additional group structure on the coefficients $\theta$, and penalizes our model through the following block-wise penalty $\sum_{g}\sqrt{\sum_{k=1}^K\sum_{j:g_j=g} \theta_{kj}^2}$. Similar to the column-wise sparse model, the block-wise sparse model does feature 
selection for all response classes at the same time, yet it may select a group of features together instead of singletons and hence relies more on the feature correlation group structure. 

Now we are ready to give the R command in $\Rcommand{Libra}$ to generate regularization paths for multinomial logistic regression.
\begin{equation*}
\Rcommand{ lb(X,y,kappa,alpha,tlist,family=``multinomial",group=FALSE,index=NA) }
\end{equation*}
We note here for the reader that the parameters $\Rcommand{alpha}$, $\Rcommand{kappa}$ and $\Rcommand{tlist}$
function the same as they do in the linear regression model, and therefore, we omit introduction of these parameters here 
but refer the reader to section~\ref{sec: linear model} for a detailed explanation of these parameters. 
Now, we are going to illustrate how the three different types of sparsity structures on parameters are implemented in R. 
To get an entry-wise sparse multinomial logistic regression, one simply sets $\Rcommand{group = FALSE}$, 
and the function $\Rcommand{lb}$ will return the solution path for this model. On the other hand, to fit a column-wise/block-wise sparse model, one needs to set $\Rcommand{group=TRUE}$ and provide the additional prior group information when possible. 

Finally, we discuss some details of the algorithmic implementation in solving the sparse multinomial logistic model. 
Similar as before, the negative log-likelihood of the multinomial model is used as the loss function: 
\begin{equation*}
L(\theta_0,\theta) = \frac{1}{n}\sum_{i=1}^n \log(\sum_{k=1}^K\exp(\theta_{k0} + x_i^T\theta_k)) - \theta_{y_i0} - x_i^T\theta_{y_i}
\end{equation*}
One can compute the derivatives of the above loss function with respect to its parameters: 
\begin{eqnarray}
\nabla_{\theta_{j0}}L(\theta_0,\theta) &= &\frac{1}{n}\sum_{i=1}^n \frac{\exp(\theta_{j0} + x_i^T\theta_j)}{\sum_{k=1}^K\exp(\theta_{k0} + x_i^T\theta_k)} - 1(y_i=j),\nonumber\\
\nabla_{\theta_j}L(\theta_0,\theta) &= &\frac{1}{n}\sum_{i=1}^n \frac{\exp(\theta_{j0} + x_i^T\theta_j)x_i}{\sum_{k=1}^K\exp(\theta_{k0} + x_i^T\theta_k)} - x_i1(y_i=j).\nonumber
\end{eqnarray}
Therefore, the computational complexity for each iteration of the Linearized Bregman Algorithm is of $O(npK)$ FLOPs. 
%
%
%
%

\renewcommand{\P}{\mathbb{P}}
\section{Graphical Model}
\label{sec: graphical model}
Undirected graphical models, also known as Markov random fields, has many applications in different fields 
including statistical physics~\cite{ising}, nature language processing~\cite{manning} and 
image analysis~\cite{hassner}, etc. Markov random field models the joint probability distribution of set random variables 
$\{X_v\}$, where the subscript $v$ belongs to some set $V$, by some undirected graph $G =(V, E)$, where 
$E \in \{0,1\}^{V \times V}$ denotes the edges among $V$ that determine the (conditional) independence between 
subsets of random variables of $\{X_v\}_{v \in V}$. In this section, we introduce three types of undirected graphical models implemented in $\Rcommand{Libra}$: Gaussian Graphical Models, Ising Models, and Potts Models. 

\subsection{Gaussian Graphic Model}
The Gaussian graphic model assumes the data $x\in \R^p$ follow the the joint normal distribution $\NN(\mu,\Theta^{-1})$, where $\Theta$ is a sparse $p$-by-$p$ inverse covariance (precision) matrix which encodes the conditional independence relations between variables, i.e. $\{x_i \perp x_j : x_{\{-i,-j\}}\} \Leftrightarrow \Theta_{ij}=0$. Note that $\theta_0$ here is the diagonal of $\Theta$ which is not penalized and the sparse parameter $\theta$ contains the off-diagonal elements. 

Graphical LASSO \cite{glasso} exploits the maximum likelihood estimate with $l_1$ regularization on $\theta$. However the gradient of Gaussian likelihood with respect to $\theta$ involves matrix inverse and is thus not a good implementation for the Linearized Bregman Algorithm. To avoid this issue, here we exploit the composite conditional likelihood as the loss function.

It is easy to calculate the distribution of $x_{j}$ conditional on $x_{-j}$ is also a normal distribution:
\begin{equation*}
x_j|x_{-j} \sim \NN\left (\mu_j - \sum_{k\neq j}\frac{\Theta_{jk}}{\Theta_{jj}}(x_k-\mu_k),\frac{1}{\Theta_{jj}}\right)
\end{equation*} 
For simplicity assume that the data is centralized, then the composite conditional likelihood becomes
\begin{equation*}
L(\Theta) = \sum_{j}^p \frac{1}{n}\sum_{i=1}^n \frac{\Theta_{jj}}{2}\left (x_{i,j} +  \sum_{k\neq j}\frac{\Theta_{jk}}{\Theta_{jj}}x_{i,k}\right)^2 - \frac{1}{2}\log\Theta_{jj}).
\end{equation*}
or equivalently, 
\begin{equation*}
L(\Theta) = \sum_{j} \frac{1}{2\Theta_{jj} }\Theta_{\cdot j}^TS\Theta_{\cdot j}- \frac{1}{2}\log(\Theta_{jj})
\end{equation*}
where $S = \frac{1}{n}\sum_{i=1}^nx_ix_i^T$ is the covariance matrix of data. Such a loss function is convex.

The corresponding gradient is defined by 
\begin{eqnarray}
\nabla_{\theta_{jj}}L(\Theta) &= &  \frac{1}{\Theta_{jj} } S_{j\cdot}\Theta_{\cdot j} - \frac{1}{2\Theta^2_{jj} }\Theta_{\cdot j}^TS\Theta_{\cdot j}- \frac{1}{2\Theta_{jj}}\nonumber\\
\nabla_{\theta_{jk}}L(\Theta) &= &  \frac{1}{\Theta_{jj} } S_{k\cdot}\Theta_{\cdot j}+\frac{1}{\Theta_{kk} } S_{j\cdot}\Theta_{\cdot k},\nonumber
\end{eqnarray}
and the computation of gradient is $O(\min(p^3,np^2))$.

The \Rcommand{Libra} command to estimate the Gaussian Graphical Model is
\begin{equation*}
\color{brown}{\tt{ggm(X,kappa,alpha,S,tlist,nt=100,trate=100)}}
\end{equation*}
where $\Rcommand{X}$ is the data matrix and if $\Rcommand{X}$ is missing, the covariance matrix $\Rcommand{S}$ should be provided. Moreover $\Rcommand{nt}$ is the number of models on path which decides the length of $\Rcommand{tlist}$ and $\Rcommand{trate}:=t_{\max}/t_{\min}$ as the scale span of $t$. Their choices are further discussed in Section~\ref{sec-detail}.

\subsubsection{Example: Journey to the West}
\begin{figure}[!h]
\centering
\includegraphics[width=0.5\textwidth]{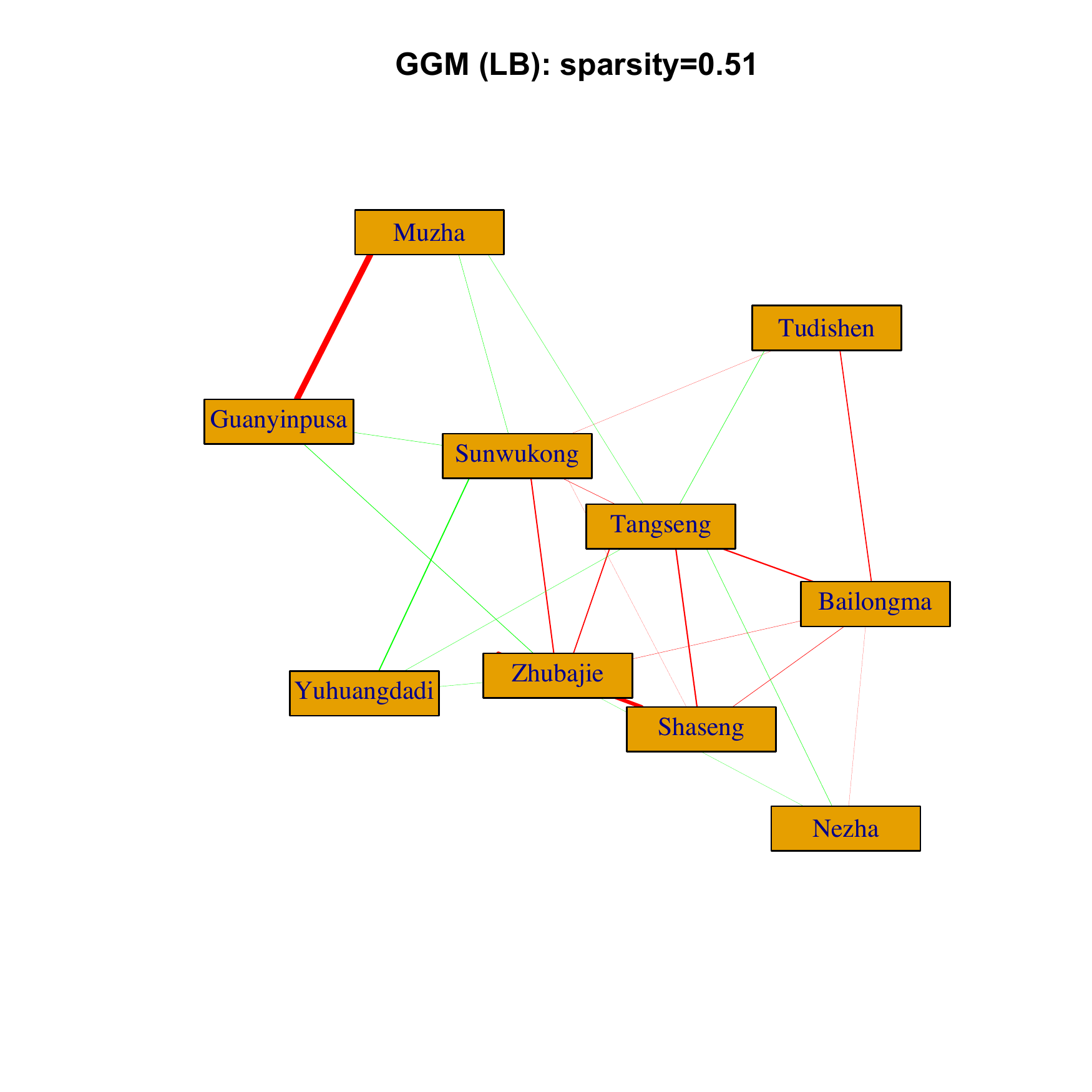}
\includegraphics[width=0.48\textwidth]{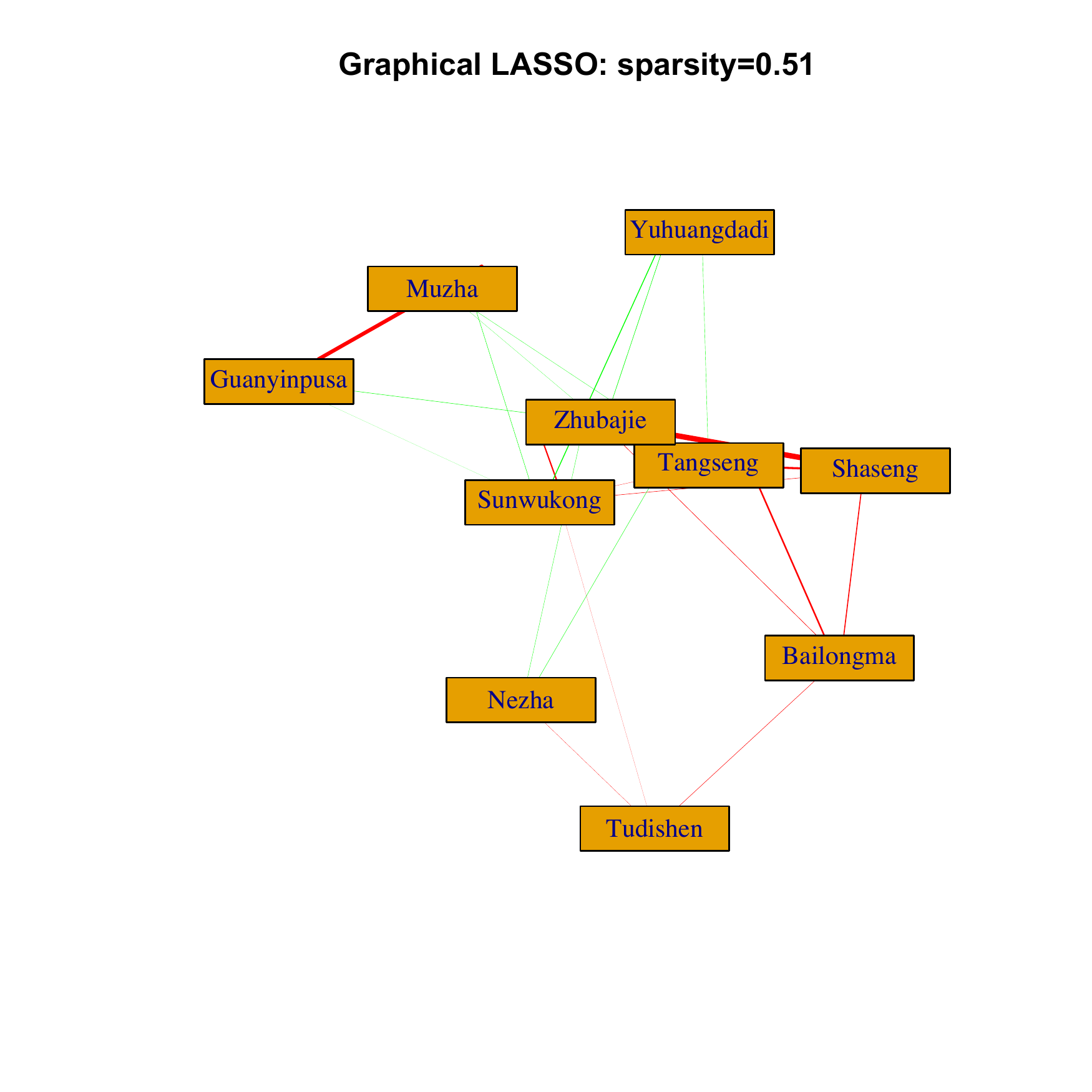}
\caption{A comparison of sparse Gaussian Graphical models returned by LB for composite conditional likelihood (left) and Graphical LASSO (right). Green for the positive coefficients and Red for the negative in the inverse covariance matrix $\Theta$. The width of edge represents the magnitude of coefficients. } \label{fig:ggm}
\end{figure}

Here we demonstrate the application of function $\Rcommand{ggm}$ to the same dataset $\Rcommand{west10}$ introduced before. We choose a particular model at sparsity level $51\%$ and plot it in Figure \ref{fig:ggm} against the outcome of Graphical LASSO implemented by R package $\Rcommand{huge}$ \cite{huge}. It can be seen that the resulting graphs bear a globally similar sparsity pattern with several distinct edges. 

\begin{lstlisting}[language=MyR,caption={}]
library(Libra)
library(igraph)
library(huge)
data(west10)

X <- as.matrix(2*west10-1);
obj = ggm(X,1,alpha = 0.01,nt=1000,trate=100)
g<-graph.adjacency(obj$path[,,720],mode="undirected",weighted=TRUE,diag=FALSE)
E(g)[E(g)$weight<0]$color<-"red"
E(g)[E(g)$weight>0]$color<-"green"
V(g)$name<-attributes(west10)$names
plot(g,vertex.shape="rectangle",vertex.size=35,vertex.label=V(g)$name,
edge.width=2*abs(E(g)$weight),main="GGM (LB): sparsity=0.51")

obj2<- huge(as.matrix(west10), method = "glasso")
obj2.select = huge.select(obj2,criterion = "ebic")
g2<-graph.adjacency(as.matrix(obj2.select$opt.icov),mode="plus",weighted=TRUE,diag=FALSE)
E(g2)[E(g2)$weight<0]$color<-"red"
E(g2)[E(g2)$weight>0]$color<-"green"
V(g2)$name<-attributes(west10)$names
plot(g2,vertex.shape="rectangle",vertex.size=35,edge.width=2*abs(E(g2)$weight),vertex.label=V(g2)$name,main="Graphical LASSO: sparsity=0.51")
\end{lstlisting}

\subsection{Ising Model}
Among many different graphical models, one important 
graphical model for \emph{binary} random variables (i.e.$X_v \in \{0, 1\}$ for any $v \in V$) is the Ising model, 
which specifies the underlying distribution on $\{X_v\}$ by the following Boltzmann distribution: 
\begin{equation*}
\P(x) = \frac{1}{Z(\theta_0, \theta)} \exp(x^T \theta_0 + \frac{1}{2} x^T \theta x), 
\end{equation*}
Here in the above equation, $\theta_0 \in \R^{|V|}$ and $\theta \in \R^{|V| \times |V|}$ are the parameters of the 
Ising model with $Z$ the normalizing function. (Z is also 
named the partition function in the literature.) Notably, the nonzero entries of $|V|$ by $|V|$ symmetric matrix 
$\theta \in \R^{|V| \times |V|}$ correspond to the edge-set $E$, which determines the dependence structure (conditional independence) between 
$\{X_v\}$. Therefore, given the data $\{x_i\}_{i=1}^n$, where $x_i \in \{0, 1\}^{|V|}$, the objective of learning 
here is to determine the support of $\theta$ (i.e., the graph structure) and estimate the strength of 
$\theta$ simultaneously (strength of dependency relationship). 

To solve this model, ~\cite{ravikumar}~\cite{xue2012} e.t.c suggest using logistic regression by observing that 
the conditional distribution of $X_v$ given all the other variables $X_{-v}$ satisfies the following logistic distribution, 
\begin{equation*}
\frac{P(X_v=1|X_{-v})}{P(X_v=0|X_{-v})} = \exp(\theta_{v0} + \theta_{v,-v}X_{-v})~~v\in V.
\end{equation*}
To fully utilize all the information from the data while keeping the symmetry of parameters, we use the following composite conditional likelihood~~\cite{xue2012}
 as our loss function in $\Rcommand{Libra}$,
 \begin{equation*}
L(\theta_0,\theta) = \sum_{v=1}^{|V|} \frac{1}{n}\sum_{i=1}^{n}\log(1+\exp(\theta_{v0} + \theta_{v,-v}x_{i,-v})) - x_{iv}(\theta_{v0} + \theta_{v,-v}x_{i,-v}), 
\end{equation*}
%
with the gradient of the above loss showing below: 
\begin{eqnarray}
\nabla_{\theta_{v0}}L(\theta_0,\theta) &= &\frac{1}{n}\sum_{i=1}^n \frac{1}{1+\exp(-\theta_{v0} - \theta_{v,-v}x_{i,-v})} - x_{iv}\nonumber\\
\nabla_{\theta_{v_1 v_2}}L(\theta_0,\theta) &= & \frac{1}{n}\sum_{i=1}^n \frac{x_{iv_2}}{1+\exp(-\theta_{v_1 0} - \theta_{v_1,-v_1}x_{i,-v_1})}+\frac{x_{iv_1}}{1+\exp(-\theta_{v_2 0} - \theta_{v_2,-v_2}x_{i,-v_2})} - 2x_{iv_1}x_{iv_2}.\nonumber
\end{eqnarray}
In fitting the Ising model, each iteration of the Linearized Bregman Algorithm requires $O(n|V|^2)$ FLOPS in general, and the 
overall time complexity for the entire solution path is $O(n|V|^2 k)$, where $k$ is the number of iterations.

The command in $\Rcommand{Libra}$ that can be used to generate the path for the Ising model is 
\begin{equation*}
\Rcommand{ising(X,kappa,alpha,tlist,responses = c(0,1),nt=100,trate=100,intercept = TRUE)}
\end{equation*}
The functions of the arguments $\Rcommand{kappa}$, $\Rcommand{alpha}$ and $\Rcommand{tilst}$ is similar to that of 
these same arguments appeared in the function calls for the linear, binomial logistic and multinomial logistic model. Hence, 
we refer the reader to section~\ref{sec: linear model} for a detailed explanations of these arguments. There are several arguments specialized 
for Ising model, i.e. $\Rcommand{nt}$ is the number of models on path which decides the length of $\Rcommand{tlist}$ and $\Rcommand{trate}:=t_{\max}/t_{\min}$ is the scale span of $t$. See section~\ref{sec-detail} for more details on these two arguments. The choice of the argument $\Rcommand{responses}$ can be either $\Rcommand{c(0,1)}$ or $\Rcommand{c(-1,1)}$. The choice $\Rcommand{c(-1,1)}$ correspond to the following model formulation, where we instead assume our data $x$ 
coming from $\{-1, 1\}$ and our distribution on data $x$ having the following specification: 
\begin{equation*}
P(x) = \frac{1}{Z} \exp(\frac{1}{2}x^Th + \frac{1}{4}x^TJ x), 
\end{equation*}
where, $h\in \R^{|V|}$ and $J \in \R^{|V| \times |V|}$. Since such model formulations appear quite often in some
scientific fields including computational physics, for convenience, we include Linearized Bregman Algorithm solvers for this type of model in our package.
For clarity, we also give the one-to-one correspondence between the two model formulations: 
\begin{eqnarray*}
x_{-1/1} & = & 2x_{0/1} -1, \\
J & = & \theta/2, \\
h & = & \theta_0 + J\mathbf{1}.
\end{eqnarray*}



\subsubsection{Example: Simulation data}
\begin{figure}[!h]
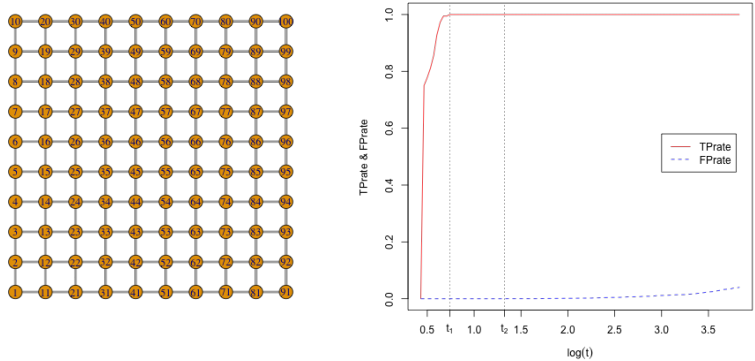

\label{fig: ising}
\includegraphics[width=0.48\textwidth]{Grid_true.png}
\includegraphics[width=0.48\textwidth]{Ising_TPFP.png}
\caption{Left: True Grid. Right: TPrate and FPrate vs. $\log(t)$. The path between $t_1$ and $t_2$ gives the correct sparsity pattern of models. }\label{fig-grid0}
\end{figure}

In this section, we give some simulation result that illustrate the performance of the Linearized Bregman Algorithm in solving the Ising model. 
In our simulation setting, we choose our sample size $n$ to be $5000$ and choose our underlying graph $G$ to be the standard $10$-by-$10$ grid (see Figure~\ref{fig-grid0}). 
We set the intercept coefficients $h$ to be $0$ for all nodes. Each entry in the interaction matrix $J_{jk}$ is set to be 
$2/2.3$ whenever $j$ and $k$ are neighbors on the $10$-by-$10$ grid or set to $0$ otherwise.  Here are the example codes that shows the simulation:
\begin{lstlisting}[language=MyR,caption={}]
library(Libra)
data(isingdata)
obj = ising(isingdata$X,10,alpha=0.1,trate=30)

TPrate <- rep(0,100)
FPrate <- rep(0,100)
for (i in 1:100){
	TPrate[i] = sum((obj$path[,,i]!=0)&(isingdata$J!=0))
	FPrate[i] = sum((obj$path[,,i]!=0)&(isingdata$J==0))
}
TPrate <- TPrate/sum(isingdata$J!=0)
FPrate <- FPrate/sum(isingdata$J==0)
tmin <- log(obj$t[min(which(TPrate==1))])
tmax <- log(obj$t[max(which(FPrate==0))])

coord = matrix(c(rep(1:10,each=10),rep(1:10,10)),ncol=2)
g<-graph.adjacency(as.matrix(isingdata$J),mode="plus",weighted=TRUE,diag=FALSE)
png(file="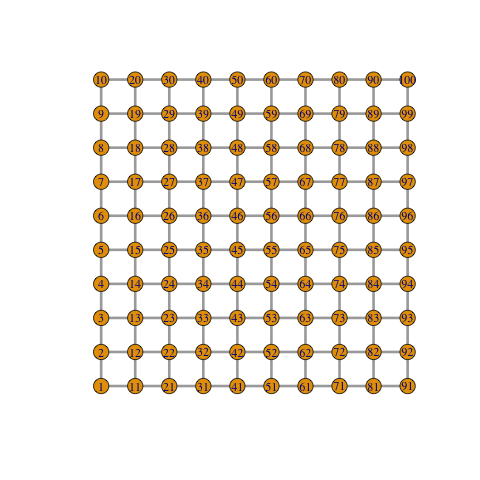", bg="transparent")
plot(g,vertex.shape="circle",vertex.size=10,edge.width=2*abs(E(g)$weight),layout=coord)
dev.off()
png(file="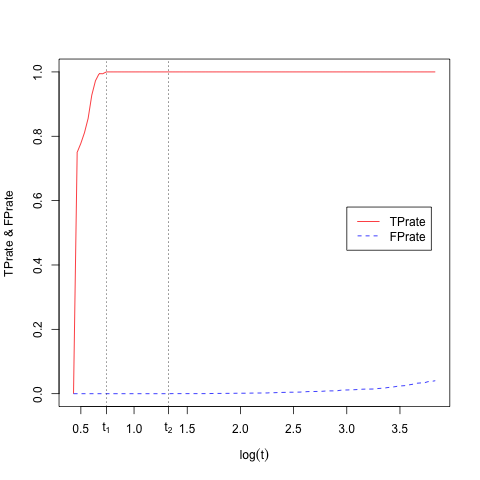", bg="transparent")
plot(log(obj$t),TPrate,col='red',type='l',lty=1,xlab=expression(log(t)),ylab='TPrate & FPrate')
lines(log(obj$t),FPrate,col='blue',type='l',lty=2)
abline(v = c(tmin,tmax),lty=3)
axis(1,at = c(tmin,tmax),labels = c(expression(t[1]),expression(t[2])))
legend(x = 3, y = 0.58, lty=1:2,col=c('red','blue'), legend=c('TPrate','FPrate'))
dev.off()
\end{lstlisting}

Figure \ref{fig-grid0} shows the True-Positive-Rate curve and False-Positive-Rate curve along the model path computed by 
$\Rcommand{ising}$. There is a segment in the LB path which gives the same sparsity pattern as the ground truth. 
 
\subsubsection{Example: Journey to the West} 
In this section, we revisit our example in section~\ref{sec: journey to the west}. In section~\ref{sec: journey to the west}, 
we analyze the social relationship between the main character~$\Rcommand{Monkey King}$ and the other 9 characters 
for the classic novel~\ref{sec: journey to the west} via a single logistic regression.  However, such analysis doesn't take into 
account the pairwise relationships between the other top 9 main characters, and hence without using the joint information 
among the other 9 characters, our estimate of social networking structure may be statistically inefficient. 
In this section, we are going to jointly estimate the social networking among all the 10 main characters simultaneously by 
applying techniques from graphical models. Note that, this can return to us a statistically more efficient estimate of the 
social networking, compared to the result coming from multiple times of single logistic regressions.  

\begin{figure}[!h]
\begin{center}
\includegraphics[width=0.60\textwidth]{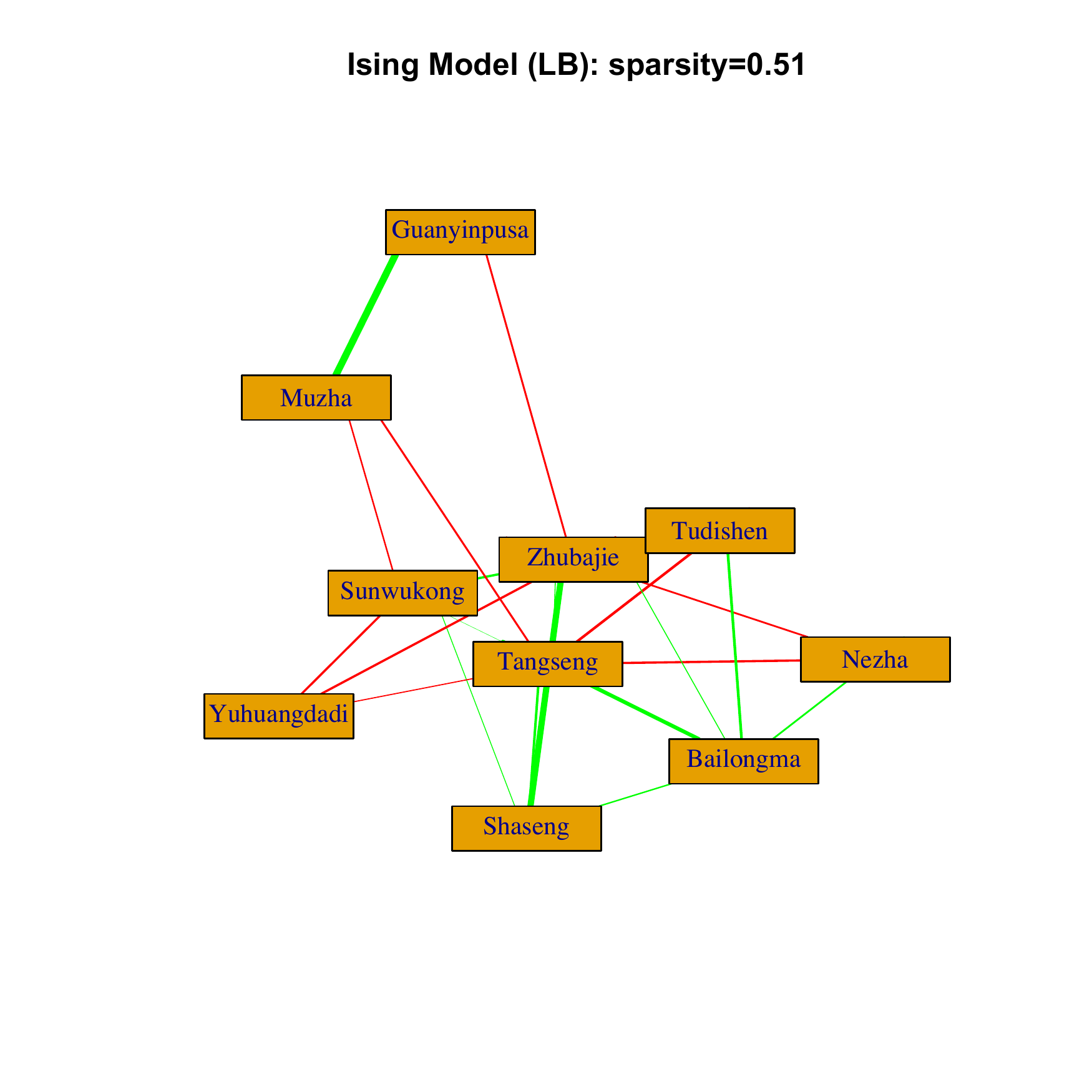}
\end{center}
\caption{An Ising model of sparsity level $51\%$ on LB path. Green edges are for positive coefficients which increase the probability of co-appearance, while red edges are for negative which drop such a probability. The width of edge represents the magnitude of coefficients. Despite that the signs of coefficients are almost opposite compared with Gaussian graphical models, the sparsity patterns in these models are qualitatively similar.} \label{fig:ising_west}
\end{figure}

Here, we first consider using Ising model to model the interaction relationships between the top 10 main characters in the 
classic novel Journey to the West. Figure \ref{fig:ising_west} shows an Ising model estimate at the same sparsity level of $51\%$ as in Figure \ref{fig:ggm}, using the command $\Rcommand{ising}$. Comparing it with Gaussian graphical models in Figure \ref{fig:ggm}, note that the color of these two types of graphs are almost opposite. This is because there is a negative sign on the exponential term in Gaussian likelihood function, which means a negative interaction coefficient actually increases the probability of co-presence in Gaussian graphical models. Up to the sign difference, the sparsity patterns in all these models are qualitatively similar. 

\begin{lstlisting}[language=MyR,caption={}]
library(Libra)
library(igraph)
data(west10)
X <- as.matrix(2*west10-1);
obj = ising(X,10,0.1,nt=1000,trate=100)

g<-graph.adjacency(obj$path[,,770],mode="undirected",weighted=TRUE)
E(g)[E(g)$weight<0]$color<-"red"
E(g)[E(g)$weight>0]$color<-"green"
V(g)$name<-attributes(west10)$names
plot(g,vertex.shape="rectangle",vertex.size=35,vertex.label=V(g)$name,edge.width=2*abs(E(g)$weight),main="Ising Model (LB): sparsity=0.51")

\end{lstlisting}

\subsubsection{Example: Dream of the Red Chamber}
\label{sec: dream of the red chamber}

Dream of the Red Chamber, often regarded as the pinnacle of Chinese fiction, is another one of the Four Great Classical Novels of Chinese Literature, composed by Cao, Xueqin for the first 80 chapters and Gao, E for the remaining 40 chapters. With a precise and detailed observation of the life and social structures typical of 18th-century society in Qing Dynasty, the novel describes a tragic romance between \Rcommand{Jia, Baoyu} and \Rcommand{Lin, Daiyu} among other conflicts. Our interest is to study the social network of interactions among the main characters. Our dataset records 375 characters who appear (`1') or do not show up (`0') in 475 events extracted from the 120 chapters. The data is collected via crowdsourcing at Peking University, and can be downloaded at the following course website:

\begin{center}
\url{http://math.stanford.edu/~yuany/course/2014.fall/}.
\end{center}

\begin{figure}[!h]
\includegraphics[width=0.48\textwidth]{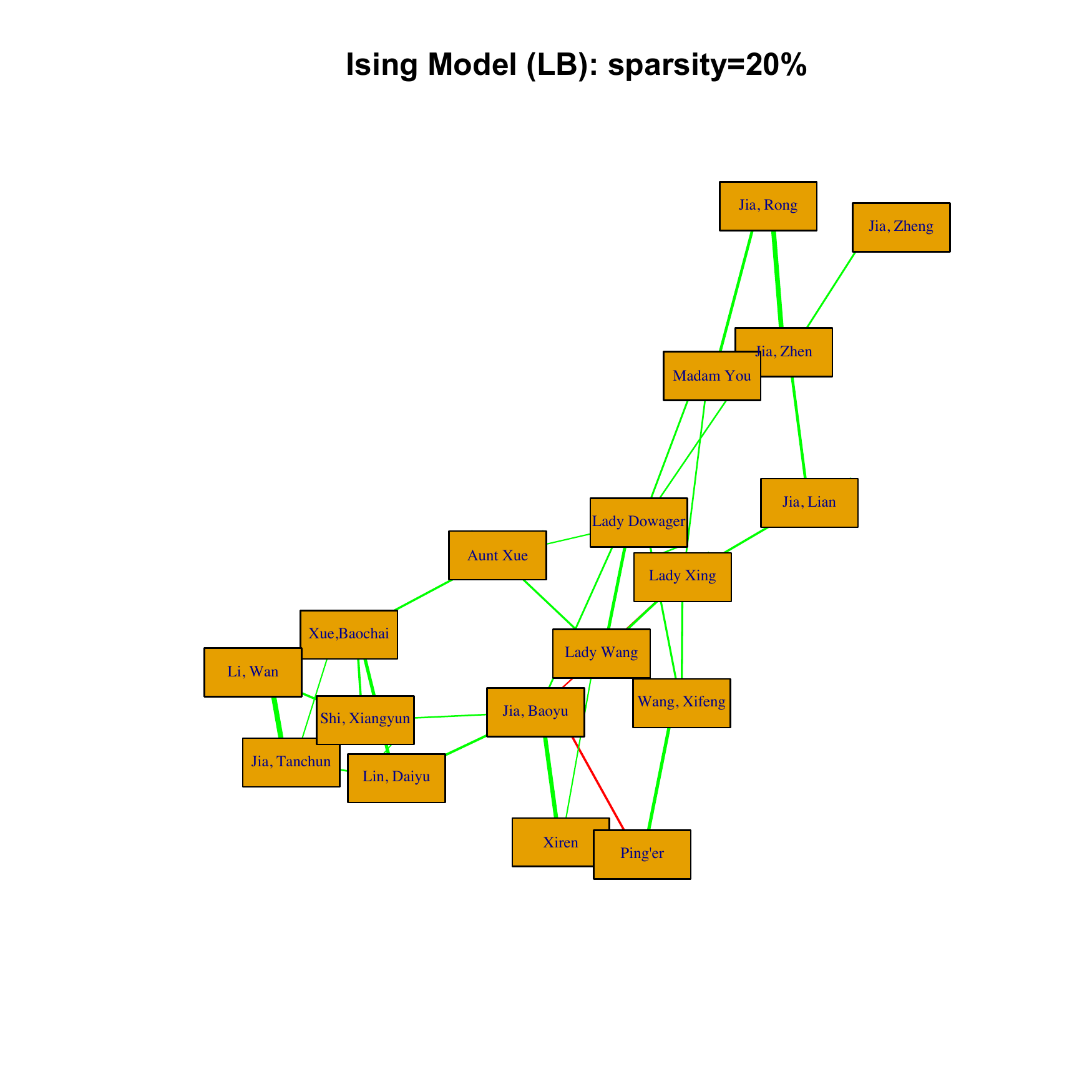}
\includegraphics[width=0.48\textwidth]{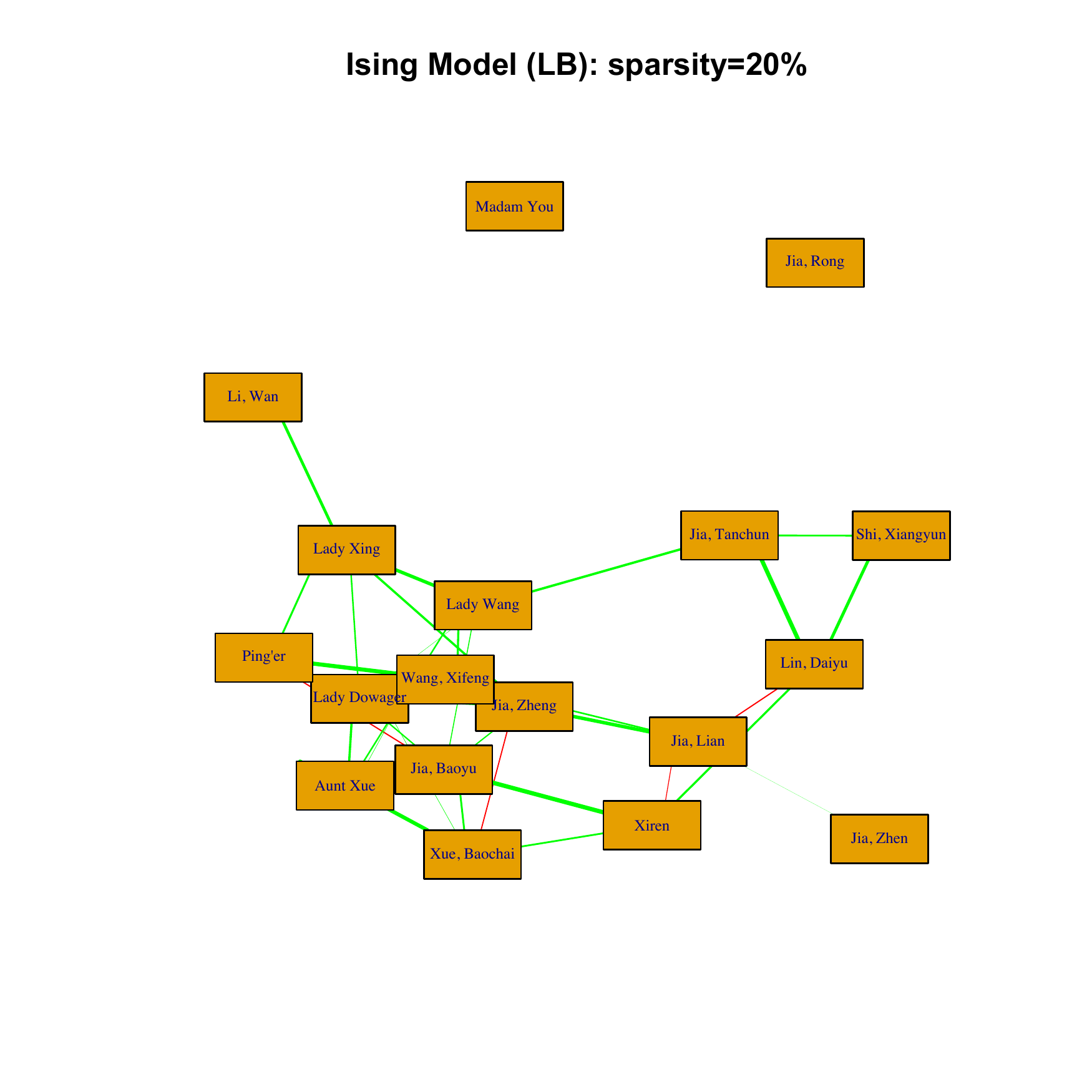}
\caption{Left: an Ising model for the first 80 chapters by Cao, Xueqin. Right: an Ising model for the remaining chapters by Gao, E. Sparsity levels are all chosen as $20\%$ on LB path. Green edges are for positive coefficients which increase the probability of co-appearance, while red edges are for negative which drop such a probability. The width of edge represents the magnitude of coefficients. Comparing the two models, one can see that \Rcommand{Jia, Baoyu} has a strong link with \Rcommand{Lin, Daiyu} in the first part, and changes the link to \Rcommand{Xue, Baochai} who becomes his wife in the second part of the novel.} \label{fig:ising_dream}
\end{figure}

The following R codes give a simple example showing how the Linearized Bregman Algorithm can be used to build up sparse Ising models from the data, focusing on the most frequently appeared 18 characters. To compare the structural difference of the first 80 chapters by Cao, Xueqin and the latter 40 chapters by Gao, E, we run $\Rcommand{ising}$ on two subsets of data to extract two Ising models shown in Figure \ref{fig:ising_dream}. The links shed light on conditional independence relations among characters learned from data. It is clear that in the first part of the novel, \Rcommand{Jia, Baoyu} has a strong connection with \Rcommand{Lin, Daiyu} and is conditional independent to another main character \Rcommand{Xue, Baochai} as Cao, Xueqin depicts; while in the second part \Rcommand{Jia, Baoyu} connects to \Rcommand{Xue, Baochai} directly and becomes conditional independent to \Rcommand{Lin, Daiyu} as Gao, E implies. Such a transition is consistent with the split of the novel. 

\begin{lstlisting}[language=MyR,caption={}]
library(Libra)
library(igraph)

load("dream.RData")
# Choose the first 80 chapters authored by Cao, Xueqin
data<-dream[dream[,1]>0,]
dim(data)
s0<-colSums(data)
# restrict to the most important characters 
data1<-data[,s0>=30]
#Eng_names <- c('Jia, Zheng','Jia, Zhen','Jia, Lian','Jia, Baoyu','Jia, Tanchun','Jia, Rong','Lady Dowager','Shi, Xiangyun','Lady Wang','Wang, Xifeng','Aunt Xue','Xue, Baochai','Lin, Daiyu','Lady Xing','Madam You','Li, Wan','Xiren','Ping\'er')
p = dim(data1)[2];
X<-as.matrix(2*as.matrix(data1[,2:p])-1);
obj = ising(X,10,0.1,nt=1000,trate=100)
sparsity=NULL
for (i in 1:1000) {sparsity[i]<-(sum(abs(obj$path[,,i])>1e-10))/(p^2-p) }

# Choose sparsity=20% at point 373
g<-graph.adjacency(obj$path[,,373],mode="undirected",weighted=TRUE)	
E(g)[E(g)$weight<0]$color<-"red"
E(g)[E(g)$weight>0]$color<-"green"
V(g)$name<-attributes(data1)$names[2:p]
plot(g,vertex.shape="rectangle",vertex.size=25,vertex.label=V(g)$name,edge.width=2*abs(E(g)$weight),vertex.label.family='STKaiti',main="Ising Model (LB): sparsity=20%")

# Choose the later 40 chapters authored by Gao, E
data<-dream[dream[,1]<1,]
data2<-data[,s0>=30]
X<-as.matrix(2*as.matrix(data2[,2:p])-1);
obj = ising(X,10,0.1,nt=1000,trate=100)
sparsity=NULL
for (i in 1:1000) {sparsity[i]<-(sum(abs(obj$path[,,i])>1e-10))/(p^2-p) }

# Choose sparsity=20% at point 344.
g<-graph.adjacency(obj$path[,,344],mode="undirected",weighted=TRUE)	
E(g)[E(g)$weight<0]$color<-"red"
E(g)[E(g)$weight>0]$color<-"green"
V(g)$name<-attributes(data2)$names[2:p]
plot(g,vertex.shape="rectangle",vertex.size=25,vertex.label=V(g)$name,edge.width=2*abs(E(g)$weight),vertex.label.family='STKaiti',main="Ising Model (LB): sparsity=20%")
\end{lstlisting}

\subsection{Potts Model}
Potts Model can be regarded as a multinomial generalization of Ising model. Each variable $x_j$ can be a multi-class variable. For simplicity  we assume $x\in\{1,2,\dots,K\}^p$, actually the class number and class name can be arbitrary. Then the model $x$ is assumed to satisfy the distribution:
\begin{equation*}
P(x) = \frac{1}{Z}\exp\left(\sum_{\substack{j=1,\dots,p\\ s=1,\dots,K}} \theta_{js,0}1(x_{j}=s) + \frac{1}{2} \sum_{\substack{j=1,\dots,p; s=1,\dots,K\\k=1,\dots,p; t=1,\dots,K}}\theta_{js,kt}1(x_j=s) 1(x_k=t)\right) 
\end{equation*}
where $Z$ is the normalization factor. The intercept coefficients $\theta_0$ is a vector of length $pK$ and the interaction coefficients $\theta$ is a $pk$-by-$pk$ symmetric matrix with zero diagonal block. So the distribution of $x_j$ conditional on the rest variables $x_{-j}$ satisfies
\begin{equation*}
P(x_j=s|x_{-j})= \frac{\exp(\theta_{js,0} + \sum_{k=1,\dots,p; t=1,\dots,K}\theta_{js,kt}1(x_k=t))}{\sum_{s=1,\dots,K}\exp(\theta_{js,0} + \sum_{k=1,\dots,p; t=1,\dots,K}\theta_{js,kt}1(x_k=t))}
\end{equation*}
which is actually a multinomial logistic distribution.

So the loss function is defined as the composite conditional likelihood:
\begin{eqnarray*}
L(\theta_0,\theta) &= &\sum_{j=1}^{p} \frac{1}{n}\sum_{i=1}^{n}\log(\sum_{s=1,\dots,K}\exp(\theta_{js,0} + \sum_{\substack{k=1,\dots,p\\ t=1,\dots,K}}\theta_{js,kt}1(x_{i,k}=t)) + \ldots \\
& & \ldots -\theta_{jx_{i,j},0} - \sum_{\substack{k=1,\dots,p\\ t=1,\dots,K}}\theta_{jx_{i,j},kt}1(x_{i,k}=t)
\end{eqnarray*}
The corresponding gradient is 
\begin{eqnarray}
\nabla_{\theta_{js,0}}L(\theta_0,\theta) &= \frac{1}{n}\sum_{i=1}^n \frac{\exp(\theta_{js,0} + \sum_{\substack{k=1,\dots,p\\ t=1,\dots,K}}\theta_{js,kt}1(x_{i,k}=t)}{\sum_{s=1,\dots,K}\exp(\theta_{js,0} + \sum_{\substack{k=1,\dots,p\\ t=1,\dots,K}}\theta_{js,kt}1(x_{i,k}=t))} - 1(x_{ij}=s)\nonumber\\
\nabla_{\theta_{js,kt}}L(\theta_0,\theta) &= \frac{1}{n}\sum_{i=1}^n \frac{1(x_{i,k}=t)\exp(\theta_{js,0} + \sum_{\substack{k=1,\dots,p\\ t=1,\dots,K}}\theta_{js,kt}1(x_{i,k}=t)}{\sum_{s=1,\dots,K}\exp(\theta_{js,0} + \sum_{\substack{k=1,\dots,p\\ t=1,\dots,K}}\theta_{js,kt}1(x_{i,k}=t))}  - 1(x_{ij}=s,x_{ik}=t)\nonumber\\
&+\frac{1(x_{i,j}=s)\exp(\theta_{kt,0} + \sum_{\substack{j=1,\dots,p\\ s=1,\dots,K}}\theta_{kt,js}1(x_{i,j}=s)}{\sum_{t=1,\dots,K}\exp(\theta_{kt,0} + \sum_{\substack{j=1,\dots,p\\ s=1,\dots,K}}\theta_{kt,js}1(x_{i,j}=s))} - 1(x_{ik}=t,x_{ij}=s)\nonumber
\end{eqnarray}
and the computation cost of gradient is $O(np^2K^2)$(or $O(np^2)$ if using sparse encoding to represent $x$).

The function to estimate the Potts model in Libra is
\begin{equation*}
\color{brown}{\tt{potts(X,kappa,alpha,tlist,nt=100,trate=100,intercept = TRUE,group=FALSE)}}
\end{equation*}
The data matrix $X$ should a matrix of size $n$-by-$p$, and each column is a class vector (The number of class for each variable can be different). If $\Rcommand{group=TRUE}$, then the group penalty is used;
\begin{equation*}
\sum_{\substack{k=1,\dots,p\\ k=1,\dots,p}} \sqrt{\sum_{\substack{s=1,\dots,K\\ t=1,\dots,K}}\theta_{js,kt}^2}.
\end{equation*}

\section{Discussions}\label{sec-detail}
In this section, we include some discussions on the choice of some universal parameters that are used throughout the $\Rcommand{Libra}$ package. 

\begin{itemize}
\item \textbf{Initialization of intercept parameter $\theta_0$:} The initialization of intercept $\theta_0$ in the Linearized Bregman Algorithm is $\theta_0^0=\arg \min_{\theta_0} L(\theta_0,0)$, not from zero. The reason for this is to avoid picking up the variables that are very relevant to the intercept term. If $\theta_0=0$ at first, then the gradient of those spurious variables close to the intercept may become very large due to the influence of intercept, such that they are much easier to be picked out. This issue is especially crucial in unbalanced sample in Ising model. When $1$ or $-1$ dominates a variable, this variable is thus very close to the intercept term and becomes a spurious variable being selected early. Fortunately, computation of $\arg \min_{\theta_0} L(\theta_0,0)$ can be done explicitly in all the examples above.
\item \textbf{Initialization of $t$:} Because the initial value of $\theta_0$ is minimal point, so the gradient of loss is always zero unless a new variable is added in. So in the package, the iteration actually begins from the first entry time 
\begin{equation*}
t_0 = \inf\{t:\theta_j(t)\ne 0,~\mbox{for some}~j\}
\end{equation*}
and $z(t_0)$ can be calculated easily because $\nabla_{\theta}L(\theta_0^0,0)$ is constant.
\item \textbf{Parameter $\Rcommand{tlist}$:} Instead of returning all the results of iteration steps, we need to return the results at a pre-decided set of $t$, $\Rcommand{tlist}$, along the path. However the Linearized Bregman Iterations only compute the value at a regular grid of time $t_0+k\alpha,k=0,1,\dots$, which may not consists a particular $t$ in $\Rcommand{tlist}$. To solve this issue, for a point $t$ in $\Rcommand{tlist}$ but not on the computed time grid, a linear interpolation of $z^k(\theta_0^{k})$ and $z^{k+1}(\theta_0^{k+1})$ is used to computed $z(t)$ or $\theta_0(t)$, $\theta(t)$ is further obtained by using $\mathbf{Shrinkage}$ on $z(t)$. Finally if $\Rcommand{tlist}$ is not specified by the user, a geometric sequence from $t_0$ to $t_0\cdot \Rcommand{trate}$ ($\Rcommand{trate}=t_{\max}/t_{\min}$) with length $\Rcommand{nt}$ (number of models on path to show) is used as the default choice $\Rcommand{tlist}$.
\end{itemize}

\begin{acknowledgement}
The authors would like to thank Chendi Huang, Stanley J. Osher, Ming Yan, and Wotao Yin for helpful discussions. The research of Jiechao Xiong and Yuan Yao was supported in part by National Basic Research Program of China: 2015CB85600 and 2012CB825501, National Natural Science Foundation of China: 61370004 and 11421110001 (A3 project), as well as grants from Baidu and Microsoft Research Asia. The research of Feng Ruan was partially supported by the E.K. Potter Stanford Graduate Fellowship.
\end{acknowledgement}

%
%
%
%
%

\bibliographystyle{abbrvnat}
\bibliography{Ref}

\end{document}